\overfullrule=0pt
\input harvmac
\def\a{{\alpha}}

\def\l{{\lambda}}
\def\lb{{\overline\lambda}}
\def\b{{\beta}}

\def\g{{\gamma}}

\def\d{{\delta}}
\def\e{{\epsilon}}

\def\half{{1\over 2}}
\def\p{{\partial}}
\def\pb{{\overline\partial}}
\def\t{{\theta}}
\def\tb{{\overline\theta}}
\def\bar{\overline}

\Title{\vbox{\hbox{IFT-P.049/2004 }}}
{\vbox{
\centerline{\bf Covariant Multiloop Superstring Amplitudes}}}
\bigskip\centerline{Nathan Berkovits\foot{e-mail: nberkovi@ift.unesp.br}}
\bigskip
\centerline{\it Instituto de F\'\i sica Te\'orica, Universidade Estadual
Paulista}
\centerline{\it Rua Pamplona 145, 01405-900, S\~ao Paulo, SP, Brasil}

\vskip .3in
\centerline{\bf Talk given at Strings 2004}
\vskip .1in
In these proceedings, the multiloop amplitude prescription using the
super-Poincar\'e invariant pure spinor formalism for the superstring is
reviewed. Unlike the RNS prescription, there is no sum over spin
structures and surface terms coming from the boundary of moduli space
can be ignored. Massless $N$-point multiloop amplitudes vanish for $N<4$, which
implies (with two mild assumptions) the perturbative finiteness of 
superstring theory. And $R^4$ terms receive no multiloop contributions 
in agreement with the Type IIB $S$-duality conjecture of Green and Gutperle.

\vskip .3in

\Date {October 2004}

\newsec{Introduction}

The computation of multiloop amplitudes in superstring theory
has many important applications such as verifying perturbative finiteness
and testing duality conjectures. Nevertheless, this subject has 
received little attention over the last fifteen years, mainly because
of difficulties in computing multiloop amplitudes using either the 
Ramond-Neveu-Schwarz (RNS) or Green-Schwarz (GS) formalism. 

In the
RNS formalism, spacetime supersymmetric amplitudes are obtained 
after summing over spin structures, which can be done explicitly only
when the number of loops and external states is small \ref\parkes
{O. Lechtenfeld and A. Parkes,
{\it On Covariant Multiloop Superstring Amplitudes},
Nucl. Phys. B332 (1990) 39.}.
Since there are divergences near the boundary of moduli space before
summing over spin structures, surface terms in the amplitude expressions
need to be treated with care \ref\verlone{E. Verlinde and H. Verlinde,
{\it Multiloop Calculations in Covariant Superstring Theory}, Phys. Lett.
B192 (1987) 95.}\ref\verltwo{
H. Verlinde, {\it The Path Integral Formulation of
Supersymmetric String Theory}, PhD Thesis, Univ. of Utrecht (1988).}
\ref\atick{J. Atick, G. Moore, and A. Sen,
{\it Catoptric Tadpoles},
Nucl. Phys. B307 (1988) 221.}
\ref\phong{E. D'Hoker and D.H. Phong, {\it Two Loop Superstrings, 
1. Main Formulas}, Phys. Lett. B529 (2002)
241, hep-th/0110247.}.
Furthermore, the complicated nature
of the Ramond vertex operator in the RNS formalism 
\ref\fms{D. Friedan, E. Martinec and S. Shenker, {\it Conformal Invariance,
Supersymmetry and String Theory}, Nucl. Phys. B271 (1986) 93.}
makes it difficult
to compute amplitudes involving external fermions or Ramond-Ramond bosons.
For these reasons, up to now, explicit multiloop computations in the RNS
formalism have been limited to four-point two-loop amplitudes involving
external Neveu-Schwarz bosons \ref\iengo{R. Iengo and C.-J. Zhu, {\it
Two Loop Computation of the Four-Particle
Amplitude in the Heterotic String}, Phys. Lett. B212 (1988) 313\semi
R. Iengo and C.-J. Zhu, {\it Explicit Modular Invariant Two-Loop Superstring
Amplitude Relevant for $R^4$}, JHEP 06 (1999) 011, hep-th/9905050.}
\phong.

In the GS formalism, spacetime supersymmetry is manifest but one needs
to fix light-cone gauge and introduce non-covariant operators at the
interaction points of the Mandelstam string diagram\ref\GS{M.B. Green
and J.H. Schwarz, {\it Superstring Interactions}, Nucl. Phys. B218 (1983)
43\semi M.B. Green and J.H. Schwarz,
{\it Superstring Field Theory}, Nucl. Phys. B243
(1984) 475.}\ref\mandelstam{S. Mandelstam,
{\it Interacting String Picture of the
Neveu-Schwarz-Ramond Model}, Nucl. Phys. B69 (1974) 77.}\ref\mandelstamtwo{
S. Mandelstam, {\it Interacting String Picture of the Fermionic String},
Prog. Theor. Phys. Suppl. 86 (1986) 163.}. Because of
complications caused by these non-covariant interaction point operators
\ref\greensite{J. Greensite and F.R. Klinkhamer, {\it Superstring 
Amplitudes and Contact Interactions}, Nucl. Phys. B304 (1988) 108\semi
M.B. Green and N. Seiberg, {\it Contact Interactions in Superstring
Theory}, Nucl. Phys. B299 (1988) 559.},
explicit amplitude expressions have been computed using the light-cone GS
formalism only for four-point tree and one-loop amplitudes \GS.

Over the past twenty years, there have been several approaches to
covariant quantization of the superstring. However, none of these
approaches were able to compute even tree-level amplitudes in a 
super-Poincar\'e covariant manner.
Four years ago, a new formalism for the superstring was proposed
\ref\superp{N. Berkovits, {\it Super-Poincar\'e Covariant Quantization of the
Superstring}, JHEP 04 (2000) 018, hep-th/0001035.}\ref\ictp{N. Berkovits,
{\it ICTP Lectures on Covariant Quantization of the
Superstring}, hep-th/0209059.}
with manifest ten-dimensional super-Poincar\'e covariance.
In conformal gauge, the worldsheet action is quadratic and physical
states are defined using a BRST operator constructed from superspace
matter variables and a pure spinor ghost variable. A super-Poincar\'e
covariant prescription was given for computing $N$-point tree amplitudes,
which was shown to coincide with the standard RNS prescription \ref
\vallilo{N. Berkovits and B.C. Vallilo, {\it
Consistency of Super-Poincar\'e Covariant Superstring
Tree Amplitudes}, JHEP 07 (2000) 015, hep-th/0004171.}\ref\relating
{N. Berkovits, {\it Relating the RNS and Pure Spinor Formalisms for the
Superstring}, JHEP 08 (2001) 026, hep-th/0104247.}.
It was also proven that the BRST cohomology reproduces the correct
superstring spectrum \ref\cohom{N. Berkovits, {\it
Cohomology in the Pure Spinor Formalism for the
Superstring}, JHEP 09 (2000) 046, hep-th/0006003\semi
N. Berkovits and O. Chand\'{\i}a,
{\it Lorentz Invariance of the Pure Spinor
BRST Cohomology for the Superstring}, 
Phys. Lett. B514 (2001) 394, hep-th/0105149.}
and that BRST invariance in a curved supergravity
background implies the low-energy superspace
equations of motion for the background superfields \ref\howeme{N. Berkovits
and P. Howe, {\it Ten-Dimensional Supergravity Constraints from the
Pure Spinor Formalism for the Superstring}
Nucl. Phys. B635 (2002) 75, hep-th/0112160.}\ref\vchan{B.C. Vallilo and O.
Chand\'{\i}a, {\it Conformal Invariance of the Pure Spinor Superstring
in a Curved Background}, JHEP 0404 (2004) 041, hep-th/0401226.}. 

Because of the pure spinor constraint satisfied by the worldsheet ghosts,
it was not obvious how to define functional integration in this formalism.
For this reason, the tree amplitude prescription in 
\superp\ relied on BRST cohomology for defining the correct normalization
of the worldsheet zero modes. Furthermore, there was no natural $b$ ghost
in this formalism, which made it difficult to define amplitudes
in a worldsheet
reparameterization-invariant manner. Because of these complications,
it was not clear how to compute loop amplitudes using this formalism and
other groups looked for ways of relaxing the pure spinor constraint
without modifying the BRST cohomology \ref\vann{P.A. Grassi, G. Policastro,
M. Porrati and P. Van Nieuwenhuizen, {\it Covariant Quantization of 
Superstrings without Pure Spinor Constraints}, JHEP 0210 (2002) 054,
hep-th/0112162\semi 
P.A. Grassi, G. Policastro and
P. Van Nieuwenhuizen, 
{\it An Introduction to the Covariant Quantization of Superstrings},
Class. Quant. Grav. 20 (2003) S395, hep-th/0302147.}
\ref\kazama{Y. Aisaka and Y. Kazama, {\it A New First Class Algebra,
Homological Perturbation and Extension of Pure Spinor Formalism for
Superstring}, JHEP 0302 (2003) 017, hep-th/0212316\semi
Y. Aisaka and Y. Kazama, {\it Operator Mapping between RNS and
Extended Pure Spinor Formalisms for Superstring}, JHEP 0308 (2003) 047,
hep-th/0305221.}
\ref\chestone{M. Chesterman, {\it Ghost Constraints
and the Covariant Quantization of the Superparticle in Ten Dimensions},
JHEP 0402 (2004) 011, hep-th/0212261\semi M. Chesterman,
{\it On the Cohomology and Inner Products of the Berkovits Superparticle
and Superstring}, hep-th/0404021.}.

Recently, it was shown 
how to perform functional integration
by defining
a Lorentz-invariant measure
for the pure spinor ghosts,
introducing appropriate ``picture-changing''
operators,
and constructing a
composite $b$ ghost in a non-zero picture.
With these three ingredients, it was straightforward to generalize
the tree amplitude prescription of \superp\
to a super-Poincar\'e covariant prescription for $N$-point 
$g$-loop amplitudes 
\ref\newloop{N. Berkovits, {\it
Multiloop Amplitudes and Vanishing Theorems using the Pure Spinor
Formalism for the Superstring}, to appear in JHEP, hep-th/0406055.}.

The need for picture-changing operators
in this formalism
is not surprising since, like
the bosonic $(\beta,\gamma)$ ghosts in the RNS formalism \fms, 
the pure
spinor ghosts are chiral bosons with worldsheet zero modes.
As in the RNS formalism, the worldsheet derivatives
of these picture-changing operators are BRST trivial so, up to possible
surface terms, the amplitudes are independent of their locations on
the worldsheet. But unlike the RNS formalism, there is no need to
sum over spin structures and there are no divergences at the boundary
of moduli space. So surface terms can be safely ignored in the
loop amplitude computations.

Although the explicit computation of arbitrary loop amplitudes is 
complicated, there are several features of the prescription
which are simpler than in the
RNS prescription. For example, there is no sum over spin structures,
no surface terms from the boundary of moduli space, and no
unphysical poles from negative-energy chiral bosons. Furthermore,
the partition functions for the matter and ghost variables cancel, 
amplitudes involving external Ramond states are no more complicated
than those involving external Neveu-Schwarz states, and
one can easily prove vanishing theorems by
counting zero modes of the fermionic superspace
variables. For example, S-duality of the Type IIB superstring implies
that $R^4$ terms in the low-energy effective action receive no 
perturbative corections above one-loop \ref\green{M.B. Green
and M. Gutperle, {\it Effects of D Instantons}, Nucl. Phys.
B498 (1997) 195, hep-th/9701093\semi M.B. Green and
P. Vanhove, {\it D-instantons, Strings and M-theory}, Phys. Lett.
B408 (1997) 122, hep-th/9704145.}.
After much effort, this was
recently verified in the RNS formalism at two-loops \iengo\phong.
Using the formalism described here, this S-duality conjecture
can be easily verified for all loops. 

Similarly, one can easily prove
the non-renormalization theorem that 
massless $N$-point multiloop amplitudes vanish whenever $N<4$.
Assuming factorization,
this non-renormalization theorem implies the absence of divergences
near the boundary of moduli space \atick
\ref\martinec{E. Martinec, {\it Nonrenormalization Theorems and
Fermionic String Finiteness}, Phys. Lett. B171 (1986) 189.}.
Note that the 
boundary of moduli space includes two types of degenerate surfaces:
surfaces where the radius $R$ of a handle shrinks to zero, and surfaces
which split into two worldsheets connected by a thin tube.
As explained in \atick, the first type of degenerate
surface does not lead to divergent amplitudes in a tachyon-free
theory since, after including
the $\log(R)$ dependence coming from integration
over the loop momenta, the amplitude integrand diverges slower than $1/R$.
The second type of degenerate surface can lead to a divergent amplitude if
there is an onshell state propagating along the thin tube between the
two worldsheets. But when all external states are on one of the two
worldsheets, vanishing of the one-point function implies the absence
of this divergence. And when all but one of the external states are on
one of the two worldsheets, vanishing of the two-point function implies
the absence of this divergence. Finally, when there are at least two
external states on each of the two worldsheets, the divergence can be
removed by analytic continuation of the external momenta \atick. Note
that vanishing of the three-point function is not required for finiteness.

So with the two mild assumptions of factorization and absence of unphysical
divergences in the interior of moduli space\foot
{In light-cone gauge, unphysical divergences in the interior
of moduli space could come from singularities between
colliding interaction points \greensite
\ref\mandfinite{S. Mandelstam,
{\it The n Loop String Amplitude: Explicit Formulas, Finiteness        
and Absence of Ambiguities},
Phys. Lett. B277 (1992) 82.}. 
In conformal gauge, there are no obvious potential
sources for these unphysical divergences in the interior of moduli space
since the amplitudes are
independent (up to surface terms)
of the locations of picture-changing operators.},
this non-renormalization theorem implies
that massless 
multiloop superstring amplitudes are finite order-by-order in
perturbation theory.
Previous attempts to prove this non-renormalization
theorem using the RNS formalism 
\martinec\
were
unsuccessful because they ignored
unphysical poles of the spacetime supersymmetry
currents \verlone\ and incorrectly assumed that the integrand
of the scattering amplitude was spacetime supersymmetric.
Using the GS formalism,
there are arguments for the non-renormalization theorem 
\ref\arguments
{A. Restuccia and J.G. Taylor, {\it Finiteness of Type II Superstring
Amplitudes}, Phys. Lett. B187 (1987) 267\semi R. Kallosh and A. Morozov,
{\it On Vanishing of Multiloop Contributions to 0,1,2,3 Point Functions
in Green-Schwarz Formalism for Heterotic String}, Phys. Lett. B207
(1988) 164.}, however, these arguments do not rule out the possibility of
unphysical divergences in the interior of moduli space from
contact term singularities between
light-cone interaction point operators \greensite.
Mandelstam was able to overcome this obstacle and prove
finiteness \mandfinite\
by combining different features of the RNS and GS formalisms.
However, the finiteness proof here is more direct than the proof of
\mandfinite\ since it is derived from a single formalism.

In section 2 of this paper, the worldsheet action and BRST operator
in the super-Poincar\'e invariant 
pure spinor formalism
of \superp\ are reviewed. 
In section 3, the three new ingredients needed for multiloop
amplitude computations are described: 
Lorentz-invariant
measure factors for the pure spinor ghosts;
picture-changing
operators; and a composite $b$ ghost in non-zero
picture.
In section 4, 
a super-Poincar\'e covariant prescription is given
for $N$-point $g$-loop amplitudes which has been shown to agree with
the RNS prescription for tree and massless four-point
one-loop amplitudes. (See \newloop\ for a more detailed version
of sections 3 and 4.)
In section 5, the counting of fermionic zero modes is
used to prove certain vanishing theorems. And in section 6,
some open questions and further applications are discussed.

\newsec{ Review of Super-Poincar\'e Invariant Pure Spinor Formalism}

\subsec{Worldsheet action}

The worldsheet variables in the Type IIB version of this formalism
include the Green-Schwarz-Siegel \ref\GScov{M.B. Green
and J.H. Schwarz, {\it Covariant Description of Superstrings},
Phys. Lett. B131 (1984) 367.}\ref\siegel{W. Siegel,
{\it Classical Superstring
Mechanics}, Nucl. Phys. B263 (1986) 93.} matter variables $(x^m,\t^\a,p_\a;
\bar\t^\a,\bar p_\a)$ for $m=0$ to 9 and $\a=1$ to 16, and the pure
spinor ghost variables $(\l^\a,w_\a;\bar\l^\a,\bar w_\a)$ where
$\l^\a$ and $\bar\l^\a$ are constrained to satisfy the pure spinor
conditions
\eqn\pure{\l^\a (\g^m)_{\a\b}\l^\b=0,\quad \bar\l^\a
(\gamma^m)_{\a\b}\bar\l^\b=0} 
for $m=0$ to 9. $(\gamma^m)_{\a\b}$ and $(\gamma^m)^{\a\b}$
are $16\times 16$ symmetric matrices which 
are the off-diagonal blocks
of the $32\times 32$ ten-dimensional $\Gamma$-matrices
and satisfy $(\g^{(m})_{\a\b} (\g^{n)})^{\b\g}=2\eta^{mn} \d_\a^\g$. 
For the 
Type IIA version
of the formalism,
the chirality of the
spinor indices on the right-moving variables is reversed, and
for the heterotic version, the right-moving variables are the same as
in the RNS formalism. 

In conformal gauge, the worldsheet action is
\eqn\action{S=\int d^2 z [-\half \p x^m\bar \p x_m -p_\a \pb\t^\a
-\bar p_\a \p\tb^\a + w_\a \pb\l^\a +\bar w_\a\p\lb^\a]}
where $\l^\a$ and $\lb^\a$ satisfy \pure.
The OPE's for the matter variables are easily computed to be
\eqn\wope{x^m(y) x^n(z) \to -\eta^{mn}\log |y-z|^2, \quad
p_\a(y)\t^\b (z) \to (y-z)^{-1} \d_\a^\b,}
however, the pure spinor constraint on $\l^\a$ prevents a direct
computation of its OPE's with $w_\a$.
As discussed in \superp,
one can solve the pure spinor constraint and express
$\l^\a$ in terms of eleven unconstrained free fields which manifestly
preserve a U(5) subgroup of the (Wick-rotated) Lorentz group. 
Although the OPE's of the unconstrained variables are not manifestly
Lorentz-covariant, 
all OPE computations involving $\l^\a$
can be expressed in a manifestly Lorentz-covariant
manner. So the non-covariant unconstrained
description of pure spinors is
useful only for verifying certain coefficients in the
Lorentz-covariant OPE's.

Because of the pure spinor constraint on $\l^\a$, the worldsheet
variables $w_\a$ contain the gauge invariance 
\eqn\gaugew{\d w_\a =\Lambda^m
(\g_m\l)_\a,}
so 5 of the 16 components of $w_\a$ can be gauged away. To preserve this
gauge invariance,
$w_\a$ can only appear in the gauge-invariant combinations 
\eqn\currents{N_{mn} = \half w_\a (\gamma_{mn})^\a{}_\b \l^\b,
\quad J= w_\a\l^\a ,}
which are the Lorentz currents and ghost current. As shown in 
\relating\ and
\cohom\ 
using either the U(5) 
or SO(8) unconstrained descriptions of pure spinors, 
$N_{mn}$ and $J$ satisfy the Lorentz-covariant OPE's
\eqn\OPE{ N_{mn}(y) \l^\a(z) \to \half (y-z)^{-1} (\gamma_{mn}\l)^\a, \quad
J(y) \l^\a(z) \to (y-z)^{-1} \l^\a,}
$$N^{kl}(y) N^{mn}(z) \to 
- 3 (y-z)^{-2}
(\eta^{n[k} \eta^{l]m}) +
(y-z)^{-1}(\eta^{m[l} N^{k]n} -
\eta^{n[l} N^{k]m} ) 
,$$
$$ J(y) J(z) \to -4 (y-z)^{-2}, \quad J(y) N^{mn}(z) \to {\rm regular}, $$
$$N_{mn}(y) T(z) \to (y-z)^{-2} N_{mn}(z) ,\quad 
J(y) T(z) \to  -8(y-z)^{-3} + (y-z)^{-2} J(z),$$
where 
\eqn\stress{T=-\half \p x^m \p x_m - p_\a \p\t^\a + w_\a \p\l^\a}
is the left-moving stress tensor. From the OPE's of \OPE, one sees
that the pure spinor condition implies that the
levels for the Lorentz and ghost currents are $-3$ and $-4$,
and that the ghost-number anomaly is $-8$. 
Note that the total Lorentz current $M^{mn}=-\half (p\g^{mn} \t) + N^{mn}$
has level $k=4-3=1$, which coincides with the level of the RNS Lorentz
current $M^{mn}=\psi^m \psi^n$. The ghost-number anomaly of $-8$ 
will be related in subsection (3.1) to the pure spinor measure factor.
Finally,
the stress tensor of \stress\
has no central charge since the $(+10-32)$ contribution
from the $(x^m,\t^\a,p_\a)$ variables is cancelled by the $+22$ contribution
from the eleven independent $(\l^\a,w_\a)$ variables. 

\subsec{BRST operator and massless vertex operators}

Physical open string states in this formalism are defined as
super-Poincar\'e covariant states of 
ghost-number $+1$ in the cohomology of the nilpotent BRST-like operator
\eqn\brst{Q = \oint \l^\a d_\a}
where 
\eqn\ddef{d_\a= p_\a -\half\g_{\a\b}^m \t^\b \p x_m -{1\over 8}
\g_{\a\b}^m \g_{m~\g\d}\t^\b\t^\g\p\t^\d}
is the supersymmetric Green-Schwarz constraint.
As shown by Siegel \siegel, $d_\a$ satisfies the OPE's 
\eqn\oped{d_\a(y) d_\b(z) \to -(y-z)^{-1} \g_{\a\b}^m \Pi_m,\quad 
d_\a(y) \Pi^m(z) \to  (y-z)^{-1} \g_{\a\b}^m \p\t^\b(z),}
$$d_\a(y) \p\t^\b(z) \to (y-z)^{-2} \d_\a^\b,\quad 
\Pi^m(y) \Pi^n(z) \to  -(y-z)^{-2} \eta^{mn},$$
where $\Pi^m = \p x^m +\half\t \g^m \p\t$
is the supersymmetric momentum and 
\eqn\defqq{
q_\a= \oint (p_\a +\half\g_{\a\b}^m \t^\b \p x_m +{1\over {24}}
\g_{\a\b}^m \g_{m~\g\d}\t^\b\t^\g\p\t^\d)}
is the supersymmetric generator satisfying
\eqn\sus{ \{q_\a,q_\b\}=
\g_{\a\b}^m\oint\p x_m, \quad [q_\a, \Pi^m(z)]=0,\quad
\{q_\a, d_\b(z)\}=0.}

To compute the massless spectrum of the open superstring,
note that
the most general vertex operator with zero conformal weight at zero momentum
and $+1$ ghost-number is 
\eqn\smax{V= \l^\a A_\a(x,\t),}
where $A_\a(x,\t)$ is a spinor superfield depending only on the worldsheet
zero modes of $x^m$ and $\t^\a$. Using the OPE that
$d_\a(y)~ f(x(z),\t(z))\to (y-z)^{-1} D_\a f$
where 
\eqn\susyd{D_\a=
{\p\over{\p\t^\a}} +\half \t^\b \g^m_{\a\b} \p_m }
is the supersymmetric
derivative, one can easily check that $QV=0$ and $\d V=Q\Lambda$
implies that $A_\a(x,\t)$ must satisfy $\l^\a\l^\b D_\a A_\b=0$
with the gauge invariance $\d A_\a=D_\a \Lambda$. But $\l^\a\l^\b
D_\a A_\b=0$ implies that 
\eqn\bianchi{D_\a A_\b + D_\b A_\a =\g^m_{\a\b} A_m}
for some vector superfield $A_m$ with the gauge transformations
\eqn\gaugeinv{\d A_\a = D_\a \Lambda, \quad \d A_m = \p_m\Lambda.}
In components, one can use \bianchi\ and \gaugeinv\ to gauge $A_\a$
and $A_m$ to the form
\eqn\opena{A_\a(x,\t)= e^{ik\cdot x}(\half a_m(\g^m\t)_\a -{1\over 3}
(\xi\g_m\t)(\g^m\t)_\a
 + ... ),}
$$A_m(x,\t) = e^{ik\cdot x}(a_m + (\xi\g^m\t) + ...),$$
where $k^2 = k^m a_m = k^m (\g_m\xi)_\a =0,$ and
$...$ involves products of $k_m$ with $a_m$ or $\xi^\a$.
So \bianchi\ and \gaugeinv\ are the equations of motion and 
gauge invariances of the ten-dimensional super-Maxwell multiplet, 
and the cohomology at ghost-number $+1$ of $Q$ correctly describes the
massless spectrum of the open superstring \ref\howe{P. Howe,
{\it Pure Spinor Lines in Superspace and Ten-Dimensional Supersymmetric
Theories}, Phys. Lett. B258 (1991) 141.}. 

As in bosonic string theory, one can obtain the integrated open string
vertex operator $\int dz U(z)$ from the unintegrated vertex operator $V$
by requiring that $QU(z)= \p V(z)$. For the massless states where
the unintegrated vertex operator is $V= \l^\a A_\a(x,\t)$, one finds that
\eqn\supermax{U = \p \t^\a A_\a (x,\t) + 
\Pi^m A_m (x,\t) + d_\a W^\a (x,\t) +\half N^{mn} {\cal F}_{mn}(x,\t)}
satisfies $QU= \p(\l^\a A_\a)$
where $A_m = {1\over 8}D_\a \g_m^{\a\b} A_\b$ is the vector gauge superfield,
$W^\b = {1\over{10}}\g_m^{\a\b} (  D_\a A^m -\p^m A_\a)$ is the
spinor superfield strength, and ${\cal F}_{mn} =
{1\over 8}D_\a (\g_{mn})^\a{}_\b W^\b = \p_{[m} A_{n]}$ is the
vector superfield strength.

\newsec{Functional Integration, Picture-Changing Operators and the $b$ Ghost}

\subsec{Measure factor for pure spinor ghosts}

As reviewed in section (2.1), the gauge invariance of \gaugew\
implies that pure spinor ghosts can only appear
through the operators 
$\l^\a$, $N_{mn}$ and $J$.
Correlation functions for the non-zero modes of these operators are easily
computed using the OPE's of \OPE. However, after integrating out the non-zero
worldsheet modes, one still has to functionally integrate over the worldsheet
zero modes. Because $\l^\a$ has zero conformal weight and satisfies
the pure spinor constraint
\eqn\puresc{\l\g^m\l=0,}
$\l^\a$ has 11 independent zero modes
on a genus $g$ surface. And because $N_{mn}$ and $J$ have $+1$ conformal
weight and are defined from gauge-invariant combinations of $w_\a$,
they have $11g$ independent zero modes on a genus $g$ surface. Note that
\puresc\ 
implies that $N_{mn}=\half (w\g_{mn}\l)$ and $J=w\l$ are related by
the equation \ref\massive
{N. Berkovits and O. Chand\'{\i}a, {\it Massive Superstring Vertex Operator in
D=10 Superspace}, JHEP
0208 (2002) 040, hep-th/0204121.} 
\eqn\Ncons{:N^{mn} \l^\a: \g_{m\a\b} -\half :J\l^\a: \g^n_{\a\b}
= 2\g^n_{\a\b}\p\l^\a}
where the normal-ordered product is defined by 
$:U^A(z)\l^\a(z): = \oint dy (y-z)^{-1} U^A(y)\l^\a(z).$ (The coefficient
of the $\p\l^\a$ term is determined by computing the double pole of the 
left-hand side of \Ncons\ with $J$.) Just as \puresc\
implies that all 16 components of $\l^\a$ can be expressed in terms
of 11 components, equation \Ncons\ implies that all
45 components of
$N^{mn}$ can be expressed in terms of $J$ and ten components of
$N^{mn}$.

Because of the constraints of \puresc\ and \Ncons, it is not immediately
obvious how to functionally integrate over the pure spinor ghosts. However,
as will be shown below, there is a natural
Lorentz-invariant measure factor for the pure spinor ghosts which can be
used to define functional integration.

A Lorentz-invariant measure factor for the $\l^\a$ zero modes can
be obtained by noting that 
\eqn\measured{(d^{11}\l)^{[\a_1\a_2 ...\a_{11}]}\equiv d\l^{\a_1} \wedge
d\l^{\a_2} \wedge ... \wedge d\l^{\a_{11}}}
satisfies the identity
\eqn\idend{ \l^\b \g^m_{\a_1 \b} 
(d^{11}\l)^{[\a_1 \a_2 ...\a_{11}]}= 0}
because $\l\g^m d\l=0$.
Using the properties of pure spinors, this implies that all 
${{16!}\over{5! 11!}}$ components
of 
$(d^{11}\l)^{[\a_1 ...\a_{11}]}$ 
are related to each other by a Lorentz-invariant
measure factor $[{\cal D}\l]$ of $+8$ ghost number which is defined by 
\eqn\mld{ 
(d^{11}\l)^{[\a_1 ...\a_{11}]}= 
[{\cal D}\l ]~~
{\cal T}_{((\b_1\b_2\b_3))}^{[\a_1 ... \a_{11}]} \l^{\b_1}\l^{\b_2}
\l^{\b_3} }
where
${\cal T}_{((\b_1\b_2\b_3))}^{[\a_1 ... \a_{11}]}$
is the unique Lorentz-invariant tensor (up to rescaling) which is symmetric
and $\g$-matrix traceless 
(i.e. $\g_m^{\b_1\b_2} {\cal T}_{((\b_1\b_2\b_3))}^{[\a_1 ...\a_{11}]}=0$)
in three lowered indices and antisymmetric in eleven
raised indices. It is defined by 
$${\cal T}_{((\b_1\b_2\b_3))}^{[\a_1 ... \a_{11}]} = 
\e^{\a_1 ...\a_{16}}(\g_{mnp})_{\a_{12}\a_{13}}
[\g^m_{\b_1\a_{14}}\g^n_{\b_2\a_{15}}\g^p_{\b_3\a_{16}} - {1\over{40}}
\g_q^{\g\d} \g^q_{(\b_1\b_2}
\g^m_{\b_3)\a_{14}}\g^n_{\g\a_{15}}\g^p_{\d\a_{16}}].$$

One can similarly construct a Lorentz-invariant measure factor for
the $N^{mn}$ and $J$ zero modes from
\eqn\measureN{(d^{11}N)^{[[m_1 n_1][m_2 n_2]...[m_{10}n_{10}]]}\equiv
dN^{[m_1 n_1]} \wedge
dN^{[m_2 n_2]} \wedge ... \wedge dN^{[m_{10} n_{10}]} \wedge dJ.}
Using the constraint of \Ncons\ and keeping $\l^\a$ fixed while
varying $N^{mn}$ and $J$, one finds that
\measureN\ satisfies the identity
\eqn\idenN{ (\l\g_{m_1})_\a 
(d^{11}N)^{[[m_1 n_1][m_2 n_2]...[m_{10}n_{10}]]}=0.}
Using the properties of pure spinors, this implies that all
${{45!}\over{10! 35!}}$ components of
$$(d^{11}N)^{[[m_1 n_1][m_2 n_2]...[m_{10}n_{10}]]}$$
are related to each
other by a Lorentz-invariant measure factor $[{\cal D}N]$
of $-8$ ghost number
which is defined by
\eqn\mlN{ 
(d^{11}N)^{[[m_1 n_1][m_2 n_2]...[m_{10}n_{10}]]} =
[{\cal D}N] }
$$\left( (\l\g^{m_1 n_1 m_2 m_3 m_4}\l)
(\l\g^{m_5 n_5 n_2 m_6 m_7}\l)
(\l\g^{m_8 n_8 n_3 n_6 m_9}\l)
(\l\g^{m_{10} n_{10} n_4 n_7 n_9}\l) + {\rm permutations}\right)$$
where the permutations are antisymmetric under the exchange of 
$m_j$ with $n_j$, and also antisymmetric under the exchange of 
$[m_j n_j]$ with $[m_k n_k]$.
Note that the index structure on the right-hand side
of \mlN\ has been chosen such the 
expression is non-vanishing after summing over the permutations.

After using the OPE's of \OPE\ to integrate out the non-zero modes of the
pure spinor ghosts on a genus $g$ surface, one will obtain an expression
\eqn\obt{{\cal A}=\langle f(\l, N_1,J_1,N_2, J_2, ..., N_g, J_g)\rangle}
which only depends on the 11 worldsheet zero modes of $\l$, and on
the 11$g$ worldsheet zero modes of $N$ and $J$. 
Using the Lorentz-invariant
measure factors defined in \mld\ and \mlN, the natural definition for
functional integration over these zero modes is
\eqn\func{{\cal A}=\int [{\cal D}\l]
[{\cal D}N_1]
[{\cal D}N_2]
...
[{\cal D}N_g]
f(\l, N_1,J_1,N_2, J_2, ..., N_g, J_g).}
Note that with this definition, 
$f(\l, N_1,J_1,N_2, J_2, ..., N_g, J_g)$ must carry ghost number 
$-8+8g$ to give a non-vanishing functional integral, which agrees with
the $-8$ ghost-number anomaly in the OPE of $J$ with $T$.
It will now be shown how the functional integral of \func\ can be explicitly
computed with the help of picture-changing operators.

\subsec{Picture-changing operators}

As is well-known from the work of Friedan-Martinec-Shenker \fms\ and
Verlinde-Verlinde \verlone\verltwo, picture-changing
operators are necessary in the RNS formalism because of the bosonic
$(\b,\g)$ ghosts. 
Since the picture-raising and picture-lowering operators involve the
delta functions $\d(\b)$ and $\d(\g)$, insertion of these operators
in loop amplitudes are needed to absorb
the zero modes of the $(\b,\g)$
ghosts on a genus $g$ surface.\foot{ 
In the RNS formalism, it is convenient to bosonize the $(\b,\g)$ ghosts
as $\b=\p\xi e^{-\phi}$ and $\gamma = \eta e^\phi$
since the spacetime supersymmetry generator involves a spin field constructed
for the negative-energy
chiral boson $\phi$. The delta functions $\d(\b)$ and $\d(\g)$
can then be expressed in terms of $\phi$ as $\d(\b)= e^\phi$ and
$\d(\g)= e^{-\phi}$. However, in the pure spinor formalism, there is
no advantage to performing such a bosonization since all operators can be
expressed directly in terms of $\l^\a$, $N^{mn}$ and $J$. Furthermore, since
functional integration over the $\phi$ chiral boson can give rise
to unphysical poles in the correlation functions, the fact that all
operators in the pure spinor formalism can be expressed in terms of
$(\l^\a,N^{mn},J)$ allows one to avoid unphysical poles in pure
spinor correlation functions.}
Up to possible surface terms, the amplitudes are independent
of the worldsheet positions of these operators since the 
worldsheet derivatives of the picture-changing operators are BRST-trivial. 
The surface terms come from pulling the BRST operator through the $b$
ghosts to give total derivatives in the worldsheet moduli. If the
correlation function diverges near the boundary of moduli space, these
surface terms can give finite contributions which need to be treated
carefully. 
As will now be shown, functional integration over the bosonic 
ghosts in the pure spinor formalism also requires picture-changing
operators with similar properties to those of the RNS formalism. 
However, since the correlation functions in this formalism do
not diverge near the boundary of moduli space, there are no
subtleties due to surface terms. 

To absorb the zero modes of $\l^\a$, $N_{mn}$ and $J$, picture-changing
operators in the pure spinor formalism will involve the delta-functions
$\d(C_\a \l^\a)$, $\d(B_{mn} N^{mn})$ and $\d(J)$ where
$C_\a$ and $B_{mn}$ are constant spinors and antisymmetric tensors.
Although these constant spinors and tensors are needed for the construction
of picture-changing operators,
it will be shown that scattering amplitudes are independent of
the choice of $C_\a$ and $B_{mn}$, so Lorentz invariance is preserved.
As will be discussed later, this Lorentz invariance can be made manifest
by integrating over all choices of $C_{\a}$ and $B_{mn}$.
Note that the use of constant spinors and tensors in picture-changing
operators is unrelated 
to the pure spinor constraint, and is necessary whenever the bosonic ghosts
are not Lorentz scalars.

As in the RNS formalism, the picture-changing
operators will be BRST-invariant with
the property that their worldsheet derivative is BRST-trivial.
A ``picture-lowering'' operator $Y_C$ with these properties is
\eqn\plo{Y_C = C_\a \t^\a \d(C_\b \l^\b)}
where $C_\a$ is any constant spinor.
Note that $Q Y_C = (C_\a\l^\a) \d(C_\b\l^\b)=0$ and
\eqn\secy{\p Y_C = (C \p\t)\d(C\l) + (C\t)(C\p\l)\d'(C\l)
= Q[ (C\p\t)(C\t) \d'(C\l)]}
where $\d'(x) \equiv {\p\over\p x}\d(x)$ is defined using the
usual rules for derivatives of delta functions, e.g. 
$x \d'(x) = -\d(x)$.

Although $Y_C$ is not spacetime-supersymmetric,
its supersymmetry variation is BRST-trivial since
\eqn\susyy{q_\a Y_C = C_\a\d(C\l) = - C_\a (C\l) \d'(C\l)=
Q[ -C_\a (C\t) \d'(C\l)].}
Similarly, $Y_C$ is not Lorentz invariant, but its Lorentz variation
is BRST-trivial since 
\eqn\lorentzy{M^{mn} Y_C = \half (C\g^{mn}\t)\d(C\l) +\half (C\t)
(C\g^{mn}\l)\d'(C\l) =
Q[ \half (C\g^{mn}\t)(C\t)\d'(C\l)].}
So different choices of $C_\a$ only change $Y_C$ by a 
BRST-trivial quantity,
and any on-shell amplitude computations involving insertions of $Y_C$
will be Lorentz invariant and spacetime supersymmetric up to possible
surface terms. 
The fact that Lorentz invariance is preserved only up
to surface terms is unrelated to the pure spinor constraint, and is
caused by the bosonic ghosts not being Lorentz scalars.

One can also construct BRST-invariant operators
involving $\d (B^{mn} N_{mn})$ and $\d (J)$
with the property that their worldsheet derivative is BRST-trivial.
These ``picture-raising'' operators will be called $Z_B$ and $Z_J$
and are defined
by
\eqn\pro{Z_B = \half B_{mn} (\l\g^{mn}d) \d(B^{pq}N_{pq}),\quad Z_J=
(\l^\a d_\a) \d(J),}
where $B_{mn}$ is a constant antisymmetric tensor. One can check
that $Q Z_B = Q Z_J=0$ and that $\p Z_B$ and $\p Z_J$
are BRST-trivial. Furthermore, different choices of $B_{mn}$ only
change $Z_B$ by a BRST-trivial quantity.

\subsec{Construction of $b$ Ghost}

To compute $g$-loop amplitudes, the usual string theory prescription
requires the insertion of  $(3g-3)$ $b$ ghosts of $-1$ ghost-number 
which satisfy
\eqn\busual{\{Q,b(u)\}= T(u)}
where $T$ is the stress tensor of \stress. After integrating $b(u)$ with
a Beltrami differential $\mu_P(u)$ for $P=1$ to $3g-3$, the BRST
variation of $b(u)$ generates a total derivative with respect to the
Teichmuller parameter $\tau_P$ associated to the Beltrami differential $\mu_P$.
But since $w_\a$
can only appear in gauge-invariant combinations of zero ghost number,
there are no operators of negative ghost number in the pure spinor
formalism, so one cannot construct such a $b$ ghost.
Nevertheless, as will now be shown, the picture-raising operator 
$$Z_B = \half B_{mn}(\l\g^{mn}d) \d(BN)$$
can be used to construct a suitable
substitute
for the $b$ ghost in non-zero picture.

Since genus $g$ amplitudes also require $10g$ insertions of $Z_B(z)$,
one can combine $(3g-3)$ insertions of $Z_B(z)$ with the desired 
insertions of the $b(u)$ ghost and look for a non-local
operator $\widetilde b_B(u,z)$ which satisfies
\eqn\bhatnew{\{Q,\widetilde b_B(u,z)\} = T(u) Z_B(z).}
Note that $Z_B$ carries $+1$
ghost-number, so $\widetilde b_B$ carries zero ghost number.
And \bhatnew\ implies that integrating $\widetilde b(u,z)$ with the
Beltrami differential $\mu_P(u)$ has the same properties as 
integrating $b(u)$ with $\mu_P(u)$ in the presence of a picture-raising
operator $Z_B(z)$. 

Using 
$$Z_B(z) = Z_B(u) + \int_u^z dv\p Z_B(v) =
Z_B(u) + \int_u^z dv \{Q, B_{pq} \p N^{pq}(v) \d(BN(v))\},$$
one can define
\eqn\learns{\widetilde b_B(u,z) = b_B(u) + T(u)
\int_u^z dv B_{pq}\p N^{pq}(v) \d(BN(v))}
where $b_B(u)$ is a local operator satisfying
\eqn\bnew{\{Q,b_B(u)\} = T(u) Z_B(u).}

The explicit formula for $b_B(u)$
satisfying \bnew\ is complicated and
was computed in \newloop\ up to some undetermined coefficients.
Ignoring Lorentz indices, $b_B$ has the form 
\eqn\bignore{b_B = B (d^2 \Pi + d N \p\t + N^2 + N \Pi^2 ) \d(BN)
+ B^2  ( d^4 + d^2 N \Pi + N^2 \Pi^2 + N^2 d \p \t) \d'(BN)}
$$+ B^3 (d^4 N + d^2 N^2 \Pi) \d''(BN)
+ B^4 (d^4 N^2) \d'''(BN).$$
For proving vanishing theorems, it will be useful to note that all
terms in $b_B$ have $+2$ conformal weight and $+4$ ``engineering'' dimension
where  $[\l,\t,x,d,N]$ are defined to carry $[0,\half,1,{3\over 2},2]$
engineering dimension and $\d(BN)$ carries $-1$
conformal weight and zero engineering dimension.

\newsec{Multiloop Amplitude Prescription}

Using the picture-changing operators and $b_B$ ghost
of section 3,
one can define a super-Poincar\'e covariant prescription
for computing $N$-point $g$-loop closed superstring scattering amplitudes as
\eqn\gloop{{\cal A} = 
\int d^2\tau_1 ...d^2\tau_{3g-3} \langle ~ |~
\prod_{P=1}^{3g-3}\int d^2 u_P \mu_P(u_P)
\widetilde b_{B_P}(u_P,z_P) }
$$
\prod_{P=3g-2}^{10g} Z_{B_P}(z_P) \prod_{R=1}^{g} Z_J(v_R)
\prod_{I=1}^{11} Y_{C_I}(y_I)~|^2 ~\prod_{T=1}^N \int d^2 t_T U_T(t_T)
~\rangle,$$
where $|~~|^2$ signifies the left-right product, 
$\tau_P$ are the Teichmuller parameters associated to the 
Beltrami differentials $\mu_P (u_P)$, and $U_T(t_T)$ are the
dimension $(1,1)$ closed string
vertex operators for the $N$ external states. 
The number of picture-lowering and picture-raising operators in
\gloop\ are appropriate for absorbing the 11 zero modes of $\l^\a$
and the $11g$ zero modes of $w_\a$, and the locations of these
picture-changing operators can be chosen arbitrarily.
The constant antisymmetric tensors $B_P^{mn}$ in $b_{B_P}$ and
$Z_{B_P}$ will be chosen such that
$B_I=B_{I+10} = ... = B_{I+10(g-1)}$ for $I=1$ to 10. In other words, there
will be ten constant antisymmetric
tensors $B_I^{mn}$, each of which appear in $g$ picture-raising
operators or $b_B$ ghosts.

When $g=1$, the prescription of \gloop\ needs to be modified for the 
usual reason that
genus-one worldsheets are invariant under constant translations, so
one of the vertex operators should be unintegrated.
The one-loop amplitude prescription is therefore
\eqn\oneloop{{\cal A} = 
\int d^2\tau \langle ~|~
\int d^2 u \mu(u)
\widetilde b_{B_1}(u,z_1) }
$$
\prod_{P=2}^{10} Z_{B_P}(z_P) Z_J(v)
\prod_{I=1}^{11} Y_{C_I}(y_I)~|^2 ~V_1(t_1)
\prod_{T=2}^N \int d^2 t_T U_T(t_T)~\rangle,$$
where $V_1(t_1)$ is the unintegrated 
closed string
vertex operator.
And when $g=0$, three of the vertex operators are unintegrated and
one uses the prescription
\eqn\zeroloop{{\cal A} = 
\langle ~|~
\prod_{I=1}^{11} Y_{C_I}(y_I)~|^2 ~V_1(t_1) V_2(t_2) V_3(t_3)
\prod_{T=4}^N \int d^2 t_T U_T(t_T)~\rangle.}

As discussed in section 3, the Lorentz variations of $\widetilde
b_{B_P}$,
$Z_{B_P}$ and $Y_{C_I}$ are BRST-trivial, so the prescription
is Lorentz-invariant up to possible surface terms. Also, all operators
are manifestly spacetime supersymmetric except for $Y_{C_I}$,
whose supersymmetry variation is BRST-trivial. In section 5, it will
be argued that surface terms can be ignored in this formalism because
of finiteness properties of the correlation functions. So the 
amplitude prescriptions of \gloop, \oneloop\ and \zeroloop\
are super-Poincar\'e covariant and
${\cal A}$ is independent of the eleven constant spinors $C_I$ and
ten constant tensors $B_P$ which appear in the picture-changing operators.
One can therefore obtain
manifestly Lorentz-covariant
expressions
from this amplitude prescription by
functionally integrating over the matter fields and
pure spinor ghosts.

As usual, the functional integration factorizes into partition
functions and correlation functions for the different worldsheet variables.
However, in the pure spinor formalism, the partition functions for the
different worldsheet
variables cancel each other out. This is easy to verify since
the partition function for the ten bosonic $x^\mu$ variables gives a factor of
$(\det \bar\p_0)^{-5}
(\det \p_0)^{-5}$ where $\p_0$ and $\bar\p_0$ are the holomorphic
and antiholomorphic derivatives acting on fields of zero conformal weight,
the 
partition function for the sixteen fermionic $(\t^\a,p_\a)$ and
$(\bar \t^\a,\bar p_\a)$ variables gives a factor of 
$(\det \bar\p_0)^{16}
(\det \p_0)^{16}$, and the partition
function for the eleven bosonic $(\l^\a,w_\a)$ and
$(\bar \l^\a,\bar w_\a)$ variables gives a factor of 
$(\det \bar\p_0)^{-11}
(\det \p_0)^{-11}$.
So to perform the functional integral, one only needs to compute the
correlation functions for the matter variables and pure spinor 
ghosts. 

As described in detail in \newloop, these correlation functions
can be computed by first separating off the zero modes from the
worldsheet variables and then using the OPE's
of \OPE\ and \oped\ 
for performing the correlation functions for the nonzero modes
of these variables. Finally, one integrates over the worldsheet
zero modes using the usual measure factors for the matter variables 
$(x^m, \t^\a, p_\a)$ and
using the Lorentz-invariant measure factors of subsection (3.1) for
the pure spinor ghost variables. 

The resulting expression for
the scattering amplitude
naively depends on the eleven constant spinors $C_I$ and ten
constant tensors $B_P$ which appear in the picture-changing operators.
However, due to Lorentz invariance of the picture-changing operators, 
one is guaranteed that this dependence on $C_I$ and $B_P$
is BRST-trivial. One can therefore integrate over all possible choices
of $C_I$ and $B_P$ and obtain a manifestly Lorentz-covariant
expression for the multiloop amplitude. As shown in \newloop,
integration over $C_I$ and $B_P$ is straightforward and the resulting covariant
expression agrees for tree amplitudes
and for massless four-point one-loop amplitudes
with the well-known RNS expression.

\newsec{Vanishing Theorems}

In this section, the amplitude prescription of section 4 will be used to
prove certain vanishing theorems
for massless closed superstring scattering amplitudes.
In subsection (5.1),
it will be proven that massless $N$-point $g$-loop amplitudes are vanishing
whenever $N<4$ and $g>0$, implying (with two mild assumptions)
the perturbative finiteness of superstring theory. 
And in subsection (5.2), it will
be proven that the low-energy limit of the four-point massless amplitude
gets no perturbative contributions above one-loop, in agreement with the
Type IIB S-duality conjecture of Green and Gutperle.

To prove these vanishing theorems, it will be useful to express the
massless closed superstring vertex
operator as the left-right product of two 
open superstring vertex operators as $V_{closed} = V_{open}\times
\overline V_{open}$ where
the closed superstring
graviton $h^{mn}$, gravitini $\psi_m^{\a}$ and $\overline \psi_m^{\a}$,
and Ramond-Ramond field strength
$F^{\a\b}$ are identified with left-right
products of the open superstring photon $a_m$ and photino $\xi^\a$ as
$$h_{mn}= a_m \overline a_n,\quad
\psi_m^{\a} = a_m \overline\xi^\a,\quad
\overline\psi_m^{\a} = \xi^\a \overline a_m,\quad
F^{\a\b} = \xi^\a\overline\xi^\b.$$
Using the unintegrated and integrated massless vertex operators of
\smax\ and \supermax, this implies that 
\eqn\closedun{V_{closed} = \l^\a\bar\l^\b A_{\a\b}(x,\t,\tb) =
e^{ik\cdot x}\l^\a A_\a(\t) \bar\l^\b \bar A_\b(\tb) \quad\quad{\rm and}}
\eqn\formvc{
U_{closed}= e^{ik\cdot x}(
\p \t^\a A_\a (\t) + 
\Pi^m A_m (\t) + d_\a W^\a (\t) +\half N^{mn} {\cal F}_{mn}(\t)) }
$$ (\bar\p \tb^\b \bar A_\b (\tb) + 
\bar\Pi^p \bar A_p (\tb) + \bar d_\b \bar W^\b (\tb) +\half \bar N^{pq} 
\bar {\cal F}_{pq}(\tb))$$
are the unintegrated and integrated massless closed superstring vertex
operators.

\subsec{Non-renormalization theorem}

In this subsection, the amplitude prescription of section 4 will be
used to prove that massless $N$-point $g$-loop amplitudes
vanish whenever $N<4$ and $g>0$.
For $N=0$, this implies vanishing of the cosmological constant; for
$N=1$, it implies absence of tadpoles; for $N=2$, it implies the mass
is not renormalized; and for $N=3$, it implies the coupling constant
is not renormalized. Using the arguments of \atick\martinec\ which were
summarized in the introduction, and
assuming factorization and the absence 
of unphysical divergences in the interior of moduli space,
these
non-renormalization theorems imply that massless superstring 
scattering amplitudes are finite order-by-order in perturbation theory.


Although
surface terms were ignored in deriving the amplitude
prescription of section 4,
it is necessary that the proof of the non-renormalization
theorem remain valid even if one includes
such surface term contributions. Otherwise, there could be divergent
surface term contributions which would invalidate the proof. For
this reason, 
one cannot assume Lorentz invariance or spacetime supersymmetry to
prove the non-renormalization theorem since the prescription of \gloop\ is
Lorentz invariant and spacetime supersymmetric only after ignoring the
surface terms. 

Fortunately, it will be possible to prove the non-renormalization theorem
using only the counting of zero modes. Since this type of argument
implies the pointwise vanishing of the integrand of the scattering
amplitude (as opposed to only implying that the integrated amplitude
vanishes), the proof remains valid if one includes the contribution
of surface terms.

On a surface of arbitrary genus, one needs 16 zero modes of $\t^\a$ and
$\tb^\a$ for the amplitude to be non-vanishing. Since the only
operators in \gloop\ containing $\t^\a$ zero modes\foot{When expressed
in terms of the free fields $(x^m,\t^\a,p_\a)$, $\Pi^m$ and $d_\a$
contain $\t$'s without derivatives which naively could contribute
$\t^\a$ zero modes. But if the supersymmetric OPE's of \oped\
are used to integrate out the non-zero worldsheet modes, the OPE's
involving $\Pi^m$ and $d_\a$ will never produce $\t^\a$ zero modes.}
are the eleven $Y_C$
picture-lowering operators and the $U_T$ vertex operators, and since
each $Y_C$ contributes a single $\t^\a$ zero mode, the $U_T$ vertex operators
must contribute at least five $\t^\a$ and five $\tb^\a$ zero modes
for the amplitude to be non-vanishing. This immediately implies that
zero-point amplitudes vanish.

For one-point amplitudes, conservation of momentum implies that
the external state must have momentum $k^m=0$. But when $k^m=0$,
the maximum number of zero modes in the vertex operator is one $\t^\a$ and
one $\tb^\a$ coming from the superfield 
$$A_{\a\b}(\t,\tb) = h_{mn} (\g^m\t)_\a (\g^n\tb)_\b.$$
All other components in the superfields appearing in the vertex operators of
\closedun\ and \formvc\
are either fermionic or involve powers of $k^m$.
So all one-point amplitudes vanish.

To prove that massless two and three-point amplitudes
vanish for non-zero $g$, one needs to count the available zero modes
of $d_\a$, as well as the zero modes of $N_{mn}$. On a genus $g$
surface, non-vanishing amplitudes require $16g$ zero modes of $d_\a$.
In addition, the number of $N_{mn}$ zero modes must be at least as large
as the number of derivatives acting on the delta functions $\d(BN)$
in the amplitude prescription. Otherwise, integration over the $N^{mn}$
zero modes will trivially vanish.

To prove the $N$-point
$g$-loop non-renormalization theorem for $N=2$ and $N=3$,
it is useful to distinguish between one-loop amplitudes and multiloop
amplitudes.
For massless $N$-point one-loop amplitudes using the prescription of \oneloop, 
there are $(N-1)$
integrated vertex operators of \formvc, each of which can either provide a
$d_\a$ zero mode or an $N_{mn}$ zero mode. So one
has at most $(N-1-M)$ $d_\a$ zero modes and $M$ $N_{mn}$ zero modes
coming from the vertex operators where $M\leq N-1$. 
Each of the nine $Z_{B_P}$ operators and 
one
$Z_J$ operator can provide a single $d_\a$ zero mode, so to get a total of
16 $d_\a$ zero modes, $b_B$ must provide at least 
\eqn\atleast{ 16- (N-1-M)-9-1 = 7-N+M }
$d_\a$ zero modes.

It is easy to verify from \bignore\
that $b_B$ can provide a maximum of four $d_\a$
zero modes, however, the terms containing four $d_\a$ zero modes also
contain $(-1)$ $N_{mn}$ zero modes where a derivative acting on $\d(BN)$
counts as a negative $N_{mn}$ zero mode. This fact can easily be derived
from the $+4$ engineering dimension of $b_B$ where 
$[\l^\a,\t^\a,x^m,d_\a, N_{mn}]$ are defined to carry engineering
dimension $[0,\half,1,{3\over 2},2]$ and $\d(BN)$ is defined to
carry zero engineering dimension. Since $(d)^4$ carries engineering dimension
$+6$, it can only appear in $b_B$ together with a term such as $\d'(BN)$
which carries engineering dimension $-2$.

So for $N\leq 3$ and $M=0$, \atleast\ implies
that the only way to obtain 16 $d_\a$ zero
modes is if $b_B$ provides at least four $d_\a$ zero modes. But in this
case, $b_B$ contains $(-1)$ $N_{mn}$ zero modes, so the amplitudes vanish
since there are not enough $N_{mn}$ zero modes to absorb the derivatives on
$\d(BN)$. And when $M>0$, the amplitude vanishes for $N\leq 3$
since one needs more than four $d_\a$ zero modes to come from $b_B$.

For multiloop amplitudes, the argument is similar, but one now has $N$
integrated vertex operators instead of $(N-1)$. So
the vertex operators can contribute a maximum of $(N-M)$ $d_\a$
zero modes and $M$ $N_{mn}$ zero modes where $M\leq N$. And each of the $7g+3$
$Z_B$ and $g$ $Z_J$ operators can provide a single $d_\a$ zero mode.
So to get a total of $16g$ $d_\a$ zero modes, the 
$(3g-3)$ $b_B$'s must provide at least
\eqn\atlt{ 16g - (N-M) - (7g+3) -g = 8g-3-N+M} 
$d_\a$ zero modes. Since $(3g-3)$ $b_B$'s
carry engineering dimension $12g-12$, $d_\a$ carries engineering dimension
$3\over 2$, and $N_{mn}$ carries engineering dimension $+2$, the
$(3g-3)$ $b_B$'s can provide a maximum of $(8g-8)$ $d_\a$ zero modes
with no derivatives of $\d(BN)$, or $(8g-8+{4\over 3}M)$ $d_\a$
zero modes with $M$ derivatives of $\d(BN)$. 
Since 
\eqn\inequal{8g-8 +{4\over 3} M < 8g-3-N+M}
whenever $M\leq N\leq 3$, there is no way for the $(3g-3)$
$b_B$'s to provide enough $d_\a$ zero modes without providing too many
derivatives of $\d(BN)$. 

So the $N$-point multiloop non-renormalization theorem has been proven
for $N\leq 3$. Note that when $N=4$, 
\eqn\nequal{8g-8 +{4\over 3} M \geq 8g-3-N+M}
if one chooses $M=3$ or $M=4$. So 
four-point multiloop amplitudes do not need to vanish. However,
as will be shown in subsection (6.4), one can prove that
the low-energy limit of these multiloop amplitudes vanish, which 
implies that the $R^4$ term in the effective action gets no 
perturbative corrections above one loop. 

\subsec{Absence of multiloop $R^4$ contributions}

Although the four-point massless amplitude is expected to be non-vanishing
at all loops, there is a conjecture based on S-duality of the Type IIB
effective action that $R^4$ terms in the low-energy effective action do
not get perturbative contributions above one-loop \green. After much effort,
this conjecture was recently verified in the RNS 
formalism at two loops \iengo\phong.
As will now be shown, the multiloop prescription of section 4 
can be easily used
to prove the validity of this S-duality conjecture at all loops.

It was proven using \nequal\
that the four-point
massless multiloop amplitude vanishes unless at least three of
the four integrated vertex operators contribute an $N_{mn}$ zero mode.
Since the only operators containing $\t$ zero modes are the eleven
picture-lowering operators and the external vertex operators, the
functional integral over $\t$ zero modes in the multiloop prescription
for the four-point amplitude gives an expression of the form
\eqn\rpr{|\int d^{16}\t (\t)^{11} (d_\a W_1^\a(\t) +\half N_{pq} 
{\cal F}^{pq}_1(\t))
\prod_{T=2}^4 N_{mn} {\cal F}^{mn}_T(\t) |^2 .}

Since the external vertex operators must contribute at least 5
$\t^\a$ and $\tb^\a$ zero modes, 
one easily sees that there is no way to produce an $|F^4|^2$ term
which would imply an $R^4$ term in the effective action. In fact,
by examining the component expansion of the ${\cal F}_{mn}(\t)$ 
and $W^\a(\t)$
superfields, one finds that the term with fewest number of spacetime
derivatives which contributes 5 $\t$'s and
5 $\tb$'s is $|(\p F)(\p F) F^2|^2$, which would imply a
$\p^4 R^4$ contribution to the low-energy effective action.

So it has been proven that there are no multiloop contributions to
$R^4$ terms (or $\p^2 R^4$ terms) in the low-energy effective action of
the superstring. It should be noted that this proof has assumed that
the correlation function over $x^m$ does not contribute
inverse powers of 
$k^m$ which could cancel momentum factors coming from the 
$\theta$ integration in \rpr.
Although the $x^m$ correlation function does contain poles as a function
of $k^m$ when the external vertex operators collide, these poles only
contribute to non-local terms in the effective action which involve
massless propagators, and are not expected to contribute to local terms in the
effective action such as the $R^4$ term.

\newsec{Conclusions}

As discussed in these proceedings, 
the super-Poincar\'e covariant prescription for multiloop superstring
amplitudes has several advantages over the RNS prescription.
There is no sum over spin structures, surface terms from the boundary
of moduli space can be ignored, and there are no unphysical poles from
a negative-energy chiral boson. Furthermore, 
the partition functions for the matter and ghost variables cancel, 
amplitudes involving external Ramond states are no more complicated
than those involving external Neveu-Schwarz states, and
one can easily prove certain vanishing theorems by
counting zero modes of the fermionic superspace
variables. 

Nevertheless, there are still some open questions concerning the 
super-Poincar\'e covariant prescription which would be useful to answer.
Since the formalism has only been defined in conformal gauge, it is not
yet clear how to derive the BRST operator and picture-changing operators
from a worldsheet reparameterization-invariant action analogous to the
Nambu-Goto action for the bosonic string. One clue may come from the
N=2 twistor-string formalism which has been shown at the classical
level to be related to the pure spinor formalism and the $b$ ghost
\ref\tonin{
M. Matone, L.
Mazzucato, I. Oda, D. Sorokin and M. Tonin, {\it The Superembedding
Origin of the Berkovits Pure Spinor Covariant Quantization of 
Superstrings}, Nucl. Phys. B639 (2002) 182, hep-th/0206104\semi
I. Oda and M. Tonin, {\it On the B Antighost in the Pure Spinor
Quantization of Superstrings}, hep-th/0409052.}.
Another important question is to show that the multiloop prescription
is unitary, possibly by proving its equivalence with a light-cone gauge
prescription.

There are many possible applications of the multiloop prescription described
here. For example, one could try to verify
duality conjectures which imply vanishing theorems for
higher-derivative $R^4$ terms \ref\vanhove
{M.B. Green, H-h. Kwon and P. Vanhove, {\it
Two Loops in Eleven Dimensions}, Phys. Rev. D61 (2000) 104010, 
hep-th/9910055.}, $R^4 H^{4g-4}$ terms \ref\mev{N. Berkovits and
C. Vafa, {\it Type IIB $R^4 H^{4g-4}$ Conjectures}, Nucl. Phys.
B533 (1998) 181, hep-th/9803145.}, and 
$F^{2n}$ terms \ref\taylor
{S. Stieberger and T.R. Taylor, {\it Nonabelian Born-Infeld Action and
Type I - Heterotic Duality 2: Nonrenormalization Theorems},
Nucl. Phys. B648 (2003) 3, hep-th/0209064.}. 
Another possible application is to generalize multiloop computations 
in
a flat ten-dimensional background to multiloop
computations in a Calabi-Yau background,
perhaps by using the hybrid formalism.
Finally, a recent exciting application of these methods has been
developed by Anguelova, Grassi and Vanhove \ref\vanhove{
L. Anguelova, P.A. Grassi and P. Vanhove, {\it Covariant One-Loop
Amplitudes in D=11}, hep-th/0408171.}
for computing covariant one-loop amplitudes in eleven
dimensions using the pure spinor version of the $d=11$ superparticle
\ref\elevenme{N. Berkovits, {\it Towards Covariant Quantization of the
Supermembrane}, JHEP 0209 (2002) 051, hep-th/0201151.}.

\vskip 15pt
{\bf Acknowledgements:} I would like to thank  
the organizers of Strings 2004 for a very successful conference,  
Sergey Cherkis, Michael Douglas, Michael Green,
Warren
Siegel, Mario Tonin, Brenno Carlini Vallilo, Pierre Vanhove,
Herman Verlinde and Edward Witten for useful discussions,
CNPq grant 300256/94-9, 
Pronex 66.2002/1998-9,
and FAPESP grant 99/12763-0
for partial financial support, and the Institute of Advanced Studies
and Funda\c{c}\~ao Instituto de F\'{\i}sica Te\'orica  
for their hospitality.

\listrefs

\end